\documentstyle[preprint,pra,floats,aps,psfig]{revtex}

\begin{document}

\title{Theory of nuclear spin conversion in ethylene}
\author{P.L.~Chapovsky\thanks{IA\&E,
         Russian Academy of Sciences, 630090 Novosibirsk, Russia; 
         e-mail: chapovsky@iae.nsk.su} and E.~Ilisca}
\address{Laboratoire de Physique Th\'{e}orique de la Mati\`{e}re 
Condens\'{e}e,\\ 
Universit\'{e} Paris 7--Denis Diderot, \\
2, Place Jussieu, 75251 Paris Cedex 05, FRANCE}
         
\date{\today}

\draft
\maketitle

\begin{abstract}
First theoretical analysis of the nuclear spin conversion in ethylene molecules
($^{13}$CCH$_{4}$) has been performed. The conversion rate was found equal
$\simeq 3\cdot 10^{-4}$~s$^{-1}$/Torr, which is in qualitative agreement
with the recently obtained experimental value. It was shown
that the ortho-para mixing in $^{13}$CCH$_{4}$ is dominated by the 
spin-rotation coupling. Mixing of only two pairs of ortho-para levels 
were found to contribute significantly to the spin conversion.
\end{abstract}

\vspace{1cm}

\section{Introduction}

Nuclear spin isomers of molecules were discovered
nearly 70 years ago when ortho and para hydrogen 
isomers were separated for the first time.  
Although it was realized already at the time of 
this discovery that many 
other symmetrical molecules have nuclear spin isomers too,
until recently almost nothing was known about isomers of molecules
heavier than hydrogen.
This happened because of the lack of practical separation methods.
The separation method based on deep cooling was applicable
to hydrogen (and deuterium) but failed in the case of 
heavier molecules. Presently a few new methods for spin isomer 
separation have been proposed and successfully tested 
which has advanced this field significantly.
Yet we are at the very early stage of this research.
The list of molecules for which the nuclear spin isomers
have been separated is rather short:
H$_{2}$ \cite{Bonhoeffer29NW}, CH$_3$F \cite{Krasnoperov84JETPL},
H$_{2}$O \cite{Konyukhov86JETPL}, CH$_{2}0$ \cite{Kern89CPL}, Li$_{2}$
\cite{He90JCP}, H$^+_3$ \cite{Uy97PRL}, and C$_2$H$_4$ \cite{Chap00CPL}.

The CH$_3$F molecules occupies a special place in this list
because it is the only polyatomic molecule for which the isomer conversion
mechanism has been identified. The CH$_3$F spin isomers appeared to be 
extremely stable surviving $\sim10^{9}$ gas collisions. Nevertheless, 
the isomer conversion was found to be based on ortho-para state 
mixing induced by tiny intramolecular hyperfine interactions and 
interruption of this mixing by collisions. This specific type of 
relaxation was proposed to refer to as quantum relaxation.
It is essential for the physics of molecular spin isomers to 
understand which processes are responsible for the spin conversion
in other molecules. 

Recently first separation of nuclear spin isomers of ethylene
molecules ($^{13}$CCH$_{4}$) has been performed \cite{Chap00CPL}. The 
conversion rate has been determined and it was shown that
the rate increases proportional to the gas pressure,
$\gamma/P\simeq 5\cdot 10^{-4}$~s$^{-1}$/Torr. An understanding of
mechanism of the conversion in ethylene needs
theoretical investigation of the process.

In the present paper we perform first theoretical analysis
of the spin conversion in ethylene molecules. The purpose is
to verify the consistency of the experimental data on isomer 
conversion in ethylene with the conversion by quantum relaxation. 
We consider the same isotope species,
$^{13}$CCH$_{4}$, for which the experiment was done \cite{Chap00CPL}. 
An essential difference of the present study from the well-understood
case of CH$_3$F molecules is that now we have to work with
an asymmetric-top molecule. This appeared to be much more 
complicated than the case of symmetric-top molecules. We would like
to point out that spin conversion in asymmetric tops were considered
theoretically previously for the CH$_{2}$O and H$_{2}$0 molecules
\cite{Curl67JCP}. 

\section{Isomer conversion by quantum relaxation}

\begin{figure}[htb]
\centerline{\psfig
{figure=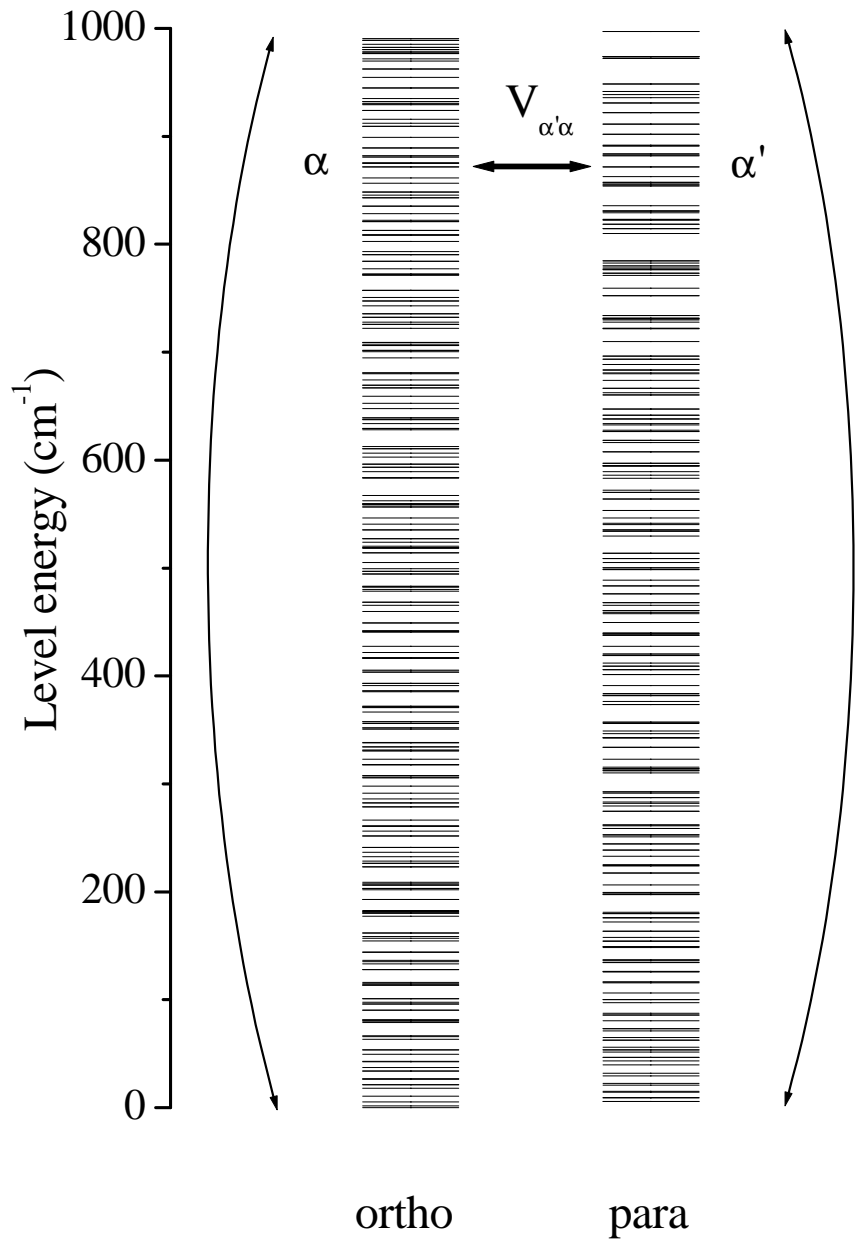,height=20cm}}
\vspace{0.5cm}
\caption{\sl The ortho and para states of the $^{13}$CCH$_{4}$
molecules. The bent lines indicate the rotational relaxation
inside the two subspaces. $V_{\alpha'\alpha}$ refers to the intramolecular 
mixing of the ortho and para states. The indicated pair of states
is the most important one for the spin conversion in $^{13}$CCH$_{4}$.}
\label{fig1}
\end{figure}

Although the spin conversion in molecules
by quantum relaxation is explained in other publications we will
describe briefly its essence here for the convenience of the reader.
At room temperature almost all molecules are situated in their
ground electronic and vibrational states. Suppose that these states
are separated into two subspaces which are the nuclear spin
ortho and para states, as is shown in Fig.~\ref{fig1}
for the particular case of the $^{13}$CCH$_{4}$ molecules. Note that
some molecules may have more than two nuclear spin isomer forms,
e.g., methane, or normal ethylene. On the other hand, number of 
molecules, e.g., H$_{2}$, CH$_3$F, and $^{13}$CCH$_{4}$ too, have 
just two type of spin species. 

The relaxation process which we are going to consider has two 
main ingredients. First, the ortho and para quantum states of 
the test molecules are not completely independent. There is 
small {\it intramolecular} perturbation, $\hat V$, which is 
able to mix the ortho and para states. This perturbation can mix, 
in general, many ortho-para level pairs. But in Fig.~\ref{fig1} 
the mixing is present just for one pair of states as will 
be the case for $^{13}$CCH$_{4}$, see below. Second, the test
molecules are embedded into an environment which is able to induce
fast relaxation inside the ortho, or para subspace, but is not
able to produce direct transitions between these subspaces.
This implies that the relevant cross-section, $\sigma(ortho|para)=0$.
The relaxation by direct transitions is the process opposite 
to the quantum relaxation.

The isomer conversion by quantum relaxation  consists in the
following. Suppose that at the instant $t=0$ a test molecule is 
placed into the ortho subspace. Due to collisions with surrounding 
particles, the test molecule starts to perform fast migration 
along rotational states inside the ortho subspace. This is the 
familiar rotational relaxation. This running up and down 
along the ladder of the ortho states  continues 
until the molecule jumps in the state $\alpha$ which is mixed
by an {\it intramolecular} perturbation with the energetically
close para state $\alpha'$. During the free flight after this collision, 
para state $\alpha'$ will be admixed to the ortho state $\alpha$. 
Consequently, the next collision can transfer the molecule
in other para states and thus localizes it inside the para subspace.
Such a mechanism of spin isomers conversion was proposed
in the theoretical work by Curl et al \cite{Curl67JCP} 
(see also \cite{Chap91PRA}).

The quantum relaxation of spin isomers can be quantitatively
described in the framework of the kinetic equation for 
density matrix \cite{Chap91PRA}. One needs to split the
molecular Hamiltonian into two parts
\begin{equation}
              \hat H = \hat H_0 + \hat V,
\label{H}              
\end{equation}
where the main part of the Hamiltonian, $\hat H_0$, has pure ortho 
and para states as the eigenstates; the perturbation $\hat V$
mixes the ortho and para states. If at initial instant the nonequilibrium
concentration of, say, ortho molecules $\delta \rho _o(t=0)$ was created, 
the system will relax then exponentially, 
$\delta \rho _o(t)=\delta \rho_o(0)e^{-\gamma t}$, with the rate 
\begin{equation}
        \gamma = \sum_{a'\in p, a\in o}
        \frac{2 \Gamma F(a'|a)}
        {\Gamma^2 + \omega^2_{a'a}}
        \left(W_p(\alpha') + W_o(\alpha)\right);\ \ 
        F(a'|a) \equiv \sum_{\nu'\in p, \nu\in o}|V_{\alpha'\alpha}|^2.
\label{gamma}        
\end{equation}
The sets of quantum numbers $\alpha'\equiv\{a',\nu'\}$ and 
$\alpha\equiv\{a,\nu\}$ consist of the degenerate quantum numbers
$\nu'$, $\nu$  and the quantum numbers $a'$, $a$ which 
determine the energy of the states. In (\ref{gamma}) $\Gamma$ is 
the decay rate of the off-diagonal density matrix element 
$\rho_{\alpha'\alpha}\ (\alpha'\in para;\ \ \alpha\in ortho)$ 
assumed here to be equal for all ortho-para level pairs; 
$\omega_{a'a}$ is the gap between the states $a'$ and $a$; 
$W_p(\alpha')$ and $W_o(\alpha)$ are the Boltzmann factors of 
the corresponding states. For the following it is 
convenient to introduce the {\it strength of mixing}, $F(a'|a)$, 
which sums the intramolecular couplings over all degenerate states.

\section{Rotational states of ethylene}

\begin{figure}[htb]
\centerline{\psfig
{figure=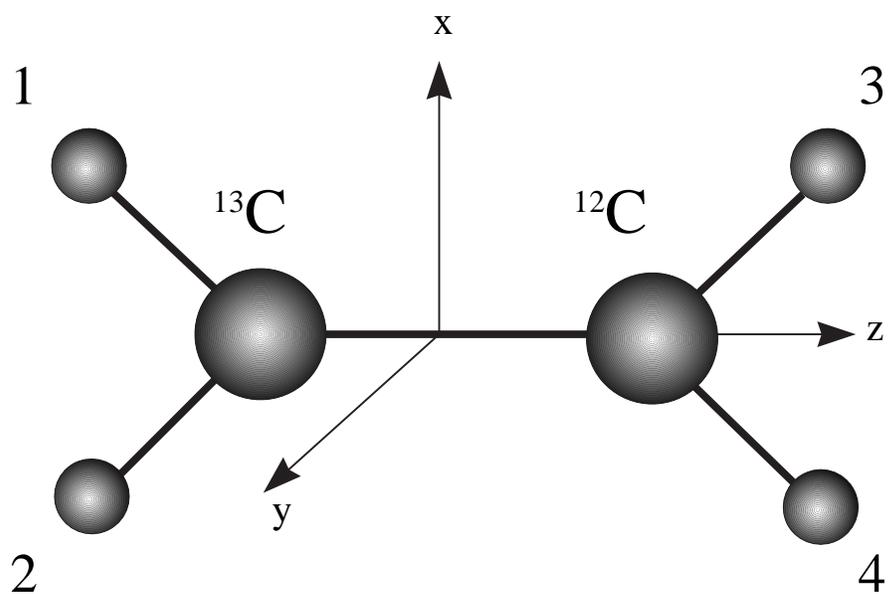,width=15cm}}
\vspace{0.5cm}
\caption{\sl Numbering of atoms in $^{13}$CCH$_{4}$ and orientation of the
molecular system of coordinates.}
\label{fig2}
\end{figure}

Ethylene is a popular object in high resolution 
infrared spectroscopy. This fortunate circumstance
made available rather accurate data on molecular parameters
and position of molecular rotational levels. The molecular
structure, numbering of hydrogen atoms and orientation of 
molecular system of coordinates are presented 
in Fig.~\ref{fig2}. $^{13}$CCH$_{4}$ is a plane,
prolate, nearly symmetric top having the symmetry point group
C$_{2v}$. The characters of the group operations and  
the irreducible representations are given in the Table~1. 
We give in the Table~1  also two isomorphic groups,
molecular symmetry group C$_{2v}(M) \cite{Bunker79}$ and
the point group D$_2$.
The bond lengths and angles which were used in our calculation
are as follows: $r_{CH}=1.087~\AA$, $r_{CC}=1.339~\AA$, the angle
$\alpha_{HCH}$=117.4$^0$ \cite{Hirota81JMS}.

The fact that the ethylene molecule is an asymmetric top  
complicates in two aspect the theoretical analysis of the isomer conversion.
First, the energy levels and wave functions for asymmetric tops 
can be found only numerically but not analytically as
is possible for symmetric-top molecules. Second, the
rotational quantum number $k$ (projection of molecular
angular momentum on the molecular symmetry axis) 
becomes an approximate quantum number in asymmetric tops. 
Consequently, one should calculate numerically more
ortho-para matrix elements than in the conversion of 
symmetric-top molecules, see, e.g., \cite{Chap91PRA}.

Rotational states of $^{13}$CCH$_{4}$ can be determined using 
the Hamiltonian of Watson  \cite{Watson68JCP} and the set of 
molecular parameters from Refs.~\cite{Vleeschouwer82JMS,Fayt99}. 
The rotational Hamiltonian up to sextic order terms has the 
form \cite{Watson68JCP}
\begin{eqnarray}
\hat H_0 & = & \frac{1}{2}(B+C){\bf J}^2+(A-\frac{1}{2}(B+C))J^2_z -
       \Delta_J({\bf J}^2)^2-\Delta_{JK}{\bf J}^2J^2_z -\Delta_K J^4_z \nonumber \\
      &&  + H_J({\bf J}^2)^3 + H_{JK}({\bf J}^2)^2J^2_z 
          + H_{KJ}{\bf J}^2J^4_z+H_KJ^6_z \nonumber \\
      && + \frac{1}{4}(B-C)F_0-\delta_J{\bf J}^2F_0-
           \delta_K {}F_2 + h_J({\bf J}^2)^2F_0
         + h_{JK}{\bf J}^2F_2 +h_KF_4 
\label{h0}
\end{eqnarray}
where ${\bf J}$, $J_x$, $J_y$, and $J_z$ are the molecular angular 
momentum operator and its projections on the molecular axes. 
The $B$, $C$, and $A$ are the parameters of a rigid top which 
characterize the rotation around $x$, $y$, and $z$ molecular axes, 
respectively (see, Fig.~\ref{fig2}). The rest of parameters account 
for the centrifugal distortion effects \cite{Watson68JCP}. 
In Eq.~(\ref{h0}) the notation was used
\begin{equation}
     F_n\equiv J^n_z(J^2_x-J^2_y) +  (J^2_x-J^2_y)J^n_z.
\label{f}
\end{equation}

The Hamiltonian (\ref{h0}) can be diagonalized for each value of $J$ 
and $M$ in the basis of the symmetric-top quantum states $|J,k,M>$
($-J\leq k \leq J$),
where $J$, $k$ and $M$ are the quantum numbers of the angular momentum,  
its projection on the molecular symmetry axis and on the
laboratory quantization axis, respectively. The Hamiltonian 
(\ref{h0}) has diagonal in $k$
matrix elements due to the first two lines and matrix elements 
having $|\Delta k|=2$. The calculations of the rotational eigen states 
can be simplified if one uses the Wang basis \cite{Landau81}
with $K=|k|$:
\begin{eqnarray}
   |\alpha,p>  & = & \frac{1}{\sqrt{2}}
           \left[|\alpha> + (-1)^{J+K+p}|\overline\alpha>\right];
           \ \ 0<K\leq J,  \nonumber \\
   |\alpha_0,p>  & = & \frac{1 + (-1)^{J+p}}{2}|\alpha_0>;
   \ \ K=0.      
\label{basis}
\end{eqnarray}
Here $p=0,1$ and the sets of quantum numbers are 
$\alpha\equiv \{J,K,M\}$; $|\overline\alpha>\equiv \{J,-K,M\}$;
$\alpha_0\equiv \{J,K=0,M\}$. 
Depending on the parity of $J$, $K$ and $p$, the states (\ref{basis}) 
generate 4 different irreducible representations of the
molecular symmetry group D$_2$ (Table~1). The molecular Hamiltonian
is full symmetric (symmetry A$_1$). Consequently, the matrix 
elements between the states of different
symmetry disappear. Thus diagonalization of the total Hamiltonian
in the basis of (\ref{basis}) is reduced to the diagonalization 
of four independent submatrices, each for the states of particular symmetry.
The rotational states of asymmetric top can be expanded 
over the basis states (\ref{basis})
\begin{equation}
     |\beta,p> = \sum_K A_K|\alpha,p>,
\label{exp}
\end{equation}
where $A_K$ stands for the expansion coefficients.
The summation index, $K$, is given explicitly in (\ref{exp}),
although $A_K$ depends on the other quantum numbers as well. All 
coefficients in the expansion (\ref{exp}) are real numbers because the 
Hamiltonian (\ref{h0}) is symmetric in the basis $|\alpha,p>$.

\begin{figure}[htb]
\centerline{\psfig
{figure=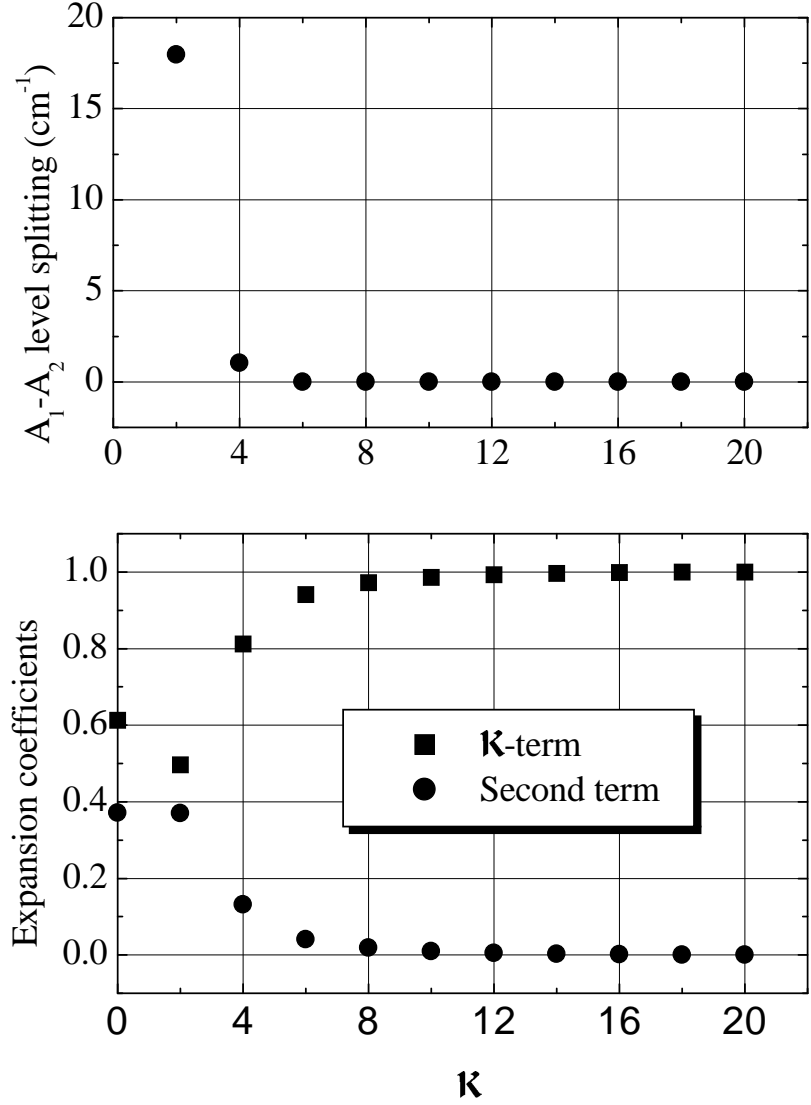,height=20cm}}
\vspace{0.5cm}
\caption{\sl Properties of asymmetric-top rotational states.
Upper panel gives splitting between the states ($0,J=20,{\cal K}$) and 
($1,J=20,{\cal K}$). Low panel gives squared values of the two terms
in the expansion (\ref{exp}) for the state ($0,J=20,{\cal K}$). 
``${\cal K}$-term'' is $A^2_{K={\cal K}}$. ``Second term'' is the second 
biggest coefficient in each expansion.}
\label{fig3}
\end{figure}

There are a few schemes for practical classification of the 
rotational states of asymmetric tops \cite{Townes55}. In this  
paper we will use the scheme which is somewhat better adapted to
the description of nuclear spin isomers.
We will designate the rotational states of asymmetric top by 
indicating $p$, $J$ and prescribing the allowed $K$ values
to the eigenstates keeping both in ascending order. For example,
the eigenstate having $p=0,\ J=20$, being the third in ascending
order will be designated by ${\cal K}=4$, because the allowed $K$ in the
expansion (\ref{exp}) are $K=0,2,4\dots$20. (Note the difference between
the two characters $K$ and ${\cal K}$). It gives unambiguous notation 
of rotational states for each of the four species A$_1$, A$_2$ 
(${\cal K}$-even) and B$_1$, B$_2$ (${\cal K}$-odd). This classification 
establishes the connection with the prolate symmetric top for which 
${\cal K}=K$. One should remember that physical meaning of the 
quantum number ${\cal K}$ is clearly limited. To illustrate this we 
consider as an example the rotational states of A$_1$ and A$_2$ 
symmetry (${\cal K}$-even) for $J=20$. The upper panel in
Fig.~\ref{fig3} shows that the energy of the rotational states is 
not determined solely by ${\cal K}$ as it would be for a rigid 
symmetric top. The graph shows the difference in energy between 
rotational states A$_1$ and A$_2$ having the same ${\cal K}$-number
but different quantum numbers $p$. The states A$_1$, ${\cal K}=0$ 
and A$_2$, ${\cal K}=0$ are omitted from the graph because
the latter state does not exist. As it is seen from the data the 
splitting is significant for low values of ${\cal K}$ and rapidly 
disappears as ${\cal K}$ increases. Note, that the A$_1$ and A$_2$ states
for a rigid symmetric-top molecule would be degenerate for all ${\cal K}$.

The low panel in Fig.~\ref{fig3} illustrates the property of the
eigen state expansions over the basis states (\ref{basis}). In this
panel the squared magnitude of the two $A_K$ coefficients in each 
eigenstate of the symmetry A$_1$ ($J=20$) is given. The first one
("${\cal K}$-term" in the Fig.~\ref{fig3}) is $A^2_K$ having $K={\cal K}$.
These coefficients appeared to be the biggest coefficients in each expansion.
This adds some physical insight to the proposed classification scheme.
The "second term" is the second biggest coefficient in the expansion.  
Again, at low ${\cal K}$ values there are more than one
significant terms in the expansion (\ref{exp}). As ${\cal K}$ grows, 
the contribution from just one
term becomes predominant. Note that the same graph
for a symmetric-top molecule would show only one significant
term, $A_{K={\cal K}}$, in the expansions of all eigenstates.

\section{Ortho and para isomers of ethylene}

The total molecular wave function is a product of spatial wave function
and spin wave function. The nuclear spin states in $^{13}$CCH$_{4}$ 
are of A$_1$ and B$_1$ species, having the statistical
weights 20 and 12, respectively. The ortho spin states 
(symmetry A$_1$) have the two hydrogen pairs (H$_1$-H$_2$ and 
H$_3$-H$_4$) either both in triplet state, or
both in singlet state. The para species (symmetry B$_1$) 
have one pair of protons in singlet state but the other pair in triplet state.
Each symmetry operation of the C$_{2v}$ group interchange in
$^{13}$CCH$_{4}$ even number of protons. Consequently,
the total wave function must be unchanged under all operations,
thus belonging to the representation A$_1$. 
In order to have the total wave function
being of species A$_1$ one has to have the spatial wave function
being of symmetry A$_1$ and B$_1$ as the spin wave functions are.
This implies that the rotational states A$_1$ and B$_1$ should be only 
positive (even in parity) but the rotational states A$_2$ and B$_2$ 
should be only negative (odd in parity). 

Summarizing this discussion we can write the total states in the 
$^{13}$CCH$_{4}$ molecules. The ortho states can be presented as
\begin{equation}
     |\mu> = |\beta,p>|i_{12},i_{34},i_C>; \ \ i_{12},i_{34}=t,s;\ \
     {\cal K}-{\text{even}}.
\label{ortho}
\end{equation}
Here $i_{12}$, $i_{34}$, and $i_C$ designate the spin states of 
the two pairs of protons and the nucleus $^{13}$C (Fig.~\ref{fig2}). 
For ortho molecules the spin states of the proton pairs should 
be either both triplet, $|t,t>$, or both singlet states, $|s,s>$.

The para states can be presented as
\begin{equation}
     |\mu'> = |\beta',p'>|i'_{12},i'_{34},i'_C>;\ \ 
     {\cal K}'-{\text{odd}}.
\label{para}
\end{equation}
For para molecules one pair of proton is in singlet, but the other 
in triplet state, thus the proton spin states are 
$|t',s'>$, or $|s',t'>$. In (\ref{ortho}), (\ref{para}) and further 
we use unprimed parameters for ortho species and primed parameters 
for para species; $p=0, 1$ indicates positive, 
or negative sign of the state, respectively. Eqs.~(\ref{ortho}),
(\ref{para})
imply that the ortho-para state mixing in ethylene needs coupling
of states having different parity of ${\cal K}$.

The relative position of ortho and para states is important for 
the calculation of the conversion rate (see Eq.~(\ref{gamma})). 
In the case of a symmetric-top molecule the 
ortho-para energy gaps can be expressed as a polynomial of $J$ 
and $K$. This gives ``accidental'' (isolated) resonances 
between the ortho and para states in symmetric top.
In the case of  an asymmetric top one has a phenomenon which can be 
called collapse of ortho and para states. It appears as a progressive 
decrease of the ortho-para energy gaps between the states of identical 
rotational momenta, $J$, as $J$ increases. This is 
illustrated in Fig.~\ref{fig4}, which shows the gaps  
between the $^{13}$CCH$_{4}$ ortho states, ${\cal K}=0$, and the para 
states, ${\cal K}'=1$, both having positive sign 
($p=0$) and the same $J$. If such sequence of 
ortho-para level pairs would exist for symmetric tops it would 
dominate the conversion. In the case of asymmetric tops the
situation is different. We will see below that this sequence of closed 
ortho-para level pairs do not produce a significant contribution to 
the isomer conversion in $^{13}$CCH$_{4}$.   

The Boltzmann factors $W_o(\alpha)$ and $W_p(\alpha')$ in 
Eq.~(\ref{gamma}) determines the relative population of 
rotational states in the ortho and para families,
\begin{equation}
     \rho_{\alpha} = \rho_o W_o(\alpha);\ \ 
     \rho_{\alpha'} = \rho_p W_p(\alpha'),
\label{Bolt}
\end{equation}
where $\rho_o$ and $\rho_p$ are the total densities of ortho 
and para molecules, respectively. The Eqs.~(\ref{Bolt}) imply 
the equilibrium distributions inside the ortho and para 
subspaces. This is fulfilled with 
high accuracy even if the ratio $\rho_o/\rho_p$
is out of equilibrium because rotational relaxation is on many 
orders of magnitude faster than the ortho-para conversion. 
The partition functions for ortho and para molecules at room 
temperature (T=295~K) are found to be equal to
\begin{equation}
     Z_{ortho}=2.66\cdot10^4;\ \ Z_{para}=1.60\cdot10^4.
\label{z}
\end{equation}
In the calculation of these partition functions we took into account
the degeneracy over $M$, nuclear spins, including also spin of
nucleus $^{13}C$, parity of states, as well as the restrictions 
imposed by quantum statistics. 

\section{Ortho-para state mixing}

In the present paper we will consider the ortho-para conversion
induced by the two hyperfine interactions, viz., magnetic
dipole-dipole interaction between the molecular nuclei (spin-spin
interaction, $\hat V_{SS}$) and the nuclear spin-rotation interaction, 
$\hat V_{SR}$. Thus the total intramolecular perturbation able 
to mix the ortho and para states is
\begin{equation}
     \hat V = \hat V_{SS} + \hat V_{SR}.
\label{v}
\end{equation}
All matrix elements of $\hat V$ are diagonal in parity $p$.

\subsection{Nuclear spin-spin coupling}

The ortho-para conversion in molecules induced by nuclear
spin-spin coupling were investigated in 
\cite{Curl67JCP,Chap91PRA} (other references see 
in the review \cite{Chap99ARPC}).
The spin-spin Hamiltonian for the two magnetic dipoles ${\bf \mu}_1$ and 
${\bf \mu}_2$ separated by the distance  $r$ has the form \cite{Landau81}
\begin{eqnarray}
     \hat V_{12} &\ =\ & 
     P_{12}{\bf T}^{(12)}{\ {}^\bullet_\bullet\ }
     \hat{\bf I}^{(1)}\hat{\bf I}^{(2)}\ ; \nonumber \\
     T_{ij}^{(12)}     &\ =\ &
     \delta _{ij}-3n_in_j\ ;\phantom{TT}P_{12}=
     \mu_1\mu_2/r^3I^{(1)}I^{(2)}h\ ,
\label{V12}     
\end{eqnarray}
where $\hat{\bf I}^{(1)}$ and $\hat{\bf I}^{(2)}$ are the spin 
operators of the particles 1 and 2, respectively; {\bf n} is
the unit vector directed along {\bf r}; $i$ and $j$ are
the Cartesian indices. The second rank tensor {\bf T}$^{(12)}$ 
in the Eq.~(\ref{V12}) represents a spatial part of the spin-spin interaction.
The second rank tensor $\hat{\bf I}^{(1)}\hat{\bf I}^{(2)}$ acts
on spin variables.

The total spin-spin interaction in $^{13}$CCH$_{4}$
is composed from the interactions between all pairs of molecular nuclei. 
One can show that the spin-spin interactions between the protons 
1-3, 2-4, 1-2, and 3-4 have the spatial part which can mix the 
quantum states only if they have $K$ numbers of the same parity. 
Consequently, these terms do not contribute to the ortho-para state mixing
in ethylene. 
The complete nuclear spin-spin interaction in $^{13}$CCH$_{4}$ 
able to mix the  ortho and para states reads
\begin{equation}
     V_{SS}= V^{(14)}_{SS}+V^{(23)}_{SS}+V^{(C1)}_{SS}+V^{(C2)}_{SS}+
     V^{(C3)}_{SS}+V^{(C4)}_{SS}.
\label{vss}
\end{equation}
Here the upper indices refer to the hydrogen nuclei in the molecule
1\dots4, and to the nucleus $^{13}$C, which has spin equal 1/2 
(see Fig.~\ref{fig2}). Calculation of the spin-spin matrix elements can 
be simplified by a few observations which we discuss 
in more detail for the perturbations $V^{(14)}_{SS}$ and $V^{(23)}_{SS}$.  
First, one can proof by applying proper symmetry operation that: 
\begin{equation}
     <\mu'|V^{(14)}_{SS}|\mu> = <\mu'|V^{(23)}_{SS}|\mu>.
\label{sym}
\end{equation}
Next, the two, out of four, matrix elements between the ortho and
para spin states vanish,
\begin{equation}
     <t',s'|V^{(14)}_{SS}|s,s> = <s',t'|V^{(14)}_{SS}|s,s> = 0,
\label{zero}
\end{equation}
because in these two cases one has the matrix elements 
of a vector (spin operator) between the states both having zero spin.
On the other hand, the remaining two matrix elements are equal 
to each other,
\begin{equation}
    <t',s'|V^{(14)}_{SS}|t,t> = <s',t'|V^{(14)}_{SS}|t,t>.
\label{non}
\end{equation}

To summarize, one can conclude that the variety of matrix elements 
for the operators $V^{(14)}_{SS}$ and $V^{(23)}_{SS}$ is reduced to one 
matrix element, e.g., $<\beta',p,t',s'|V^{(14)}_{SS}|\beta,p,t,t>$. 
Obviously, the carbon spin state, which is omitted in this expression 
for simplicity, should be unchanged in this matrix element. 

Further, we write this matrix element using an expansion
over symmetric-top states (\ref{exp})
\begin{equation}
     <\beta',p,t',s'|V^{(14)}_{SS}|\beta,p,t,t> = <t',s'|
     \left[\sum_{K',K}A'_{K'}A_K 
     <\alpha',p|V^{(14)}_{SS}|\alpha,p>\right]|t,t>.    
\label{exp1}
\end{equation}
This expression reduces the calculation of the spin-spin matrix 
elements in asymmetric tops to the calculation of symmetric-top 
matrix elements. Solution for the latter can be found 
in \cite{Chap91PRA}, which allows to express the strength of 
mixing in ethylene by $\hat V_{SS}$ as
\begin{eqnarray}
   F_{SS}(a'|a) & = & (2J'+1)(2J+1){\cal T}^2
                \Bigg|\sum_{K>0,q}qA'_{K+q}A_K 
                \left(\begin{array}{ccr}
                    J' & 2 & J \\ 
                   -K-q & q & K
                \end{array}\right)        \nonumber \\
          &  & +\frac{1+(-1)^{J+p}}{\sqrt{2}}A'_1A_0
                \left(\begin{array}{rcr}
                    J' & 2 & J \\ 
                   -1  & 1 & 0    
                \end{array}\right)\Bigg|^2.       
\label{fss}  
\end{eqnarray}
Here $q=\pm1$; (:\ :\ :) stands for the 3j-symbol and the notation 
was used
\begin{equation}
     {\cal T}^2 = 2|P_{14}{\cal T}^{(14)}_{2,1}|^2 +
                        2|P_{C1}{\cal T}^{(C1)}_{2,1}|^2 +
                        2|P_{C3}{\cal T}^{(C3)}_{2,1}|^2;\ \ 
     {\cal T} = 46.5~{\text{kHz}},
\label{T}
\end{equation}
where $P$-factors are equivalent to the similar factor in (\ref{V12});
${\cal T}^{(mn)}_{2,1}$ are the spherical components of the corresponding
{\bf T}-tensor calculated in the molecular frame. The numerical value of
${\cal T}$ was found using the molecular structure from
\cite{Volkenshtein72}.
The selection rules for the spin-spin mixing in ethylene read
\begin{equation}
     \Delta p = 0;\ \ |\Delta J| \leq 2,
\label{selSS}
\end{equation}
and in addition, parity of ${\cal K}'$ and ${\cal K}$ is opposite.
           
\subsection{Spin-rotation coupling}

Ortho-para conversion in molecules induced by spin-rotation coupling were
studied in \cite{Curl67JCP,Guskov95JETP,Ilisca98PRA} (more references 
can be found in \cite{Chap99ARPC}). Nuclear spin-rotation coupling
in molecules is due to magnetic fields produced by the molecular 
electrical currents. The spin-rotation perturbation can be presented 
as \cite{Gunther-Mohr54PR,Townes55,Ilisca98PRA}
\begin{equation}
     \hat V_{SR} = \frac{1}{2}
     \left(\sum_i \hat {\bf I}^{(i)}\bullet {\bf C}^{(i)} 
     \bullet \hat{\bf J} + h.c.\right);\ \ i=1,2,3,4.
\label{vsr}
\end{equation}
For the spin-rotation perturbation relevant to the ortho-para mixing, 
the index $i$ should refer only to the hydrogen nuclei.

The calculation of the second rank spin-rotation tensor {\bf C} is rather
complicated problem which is not solved yet completely. We will use the 
following estimation of {\bf C}. First we note that contribution
to the spin-rotation coupling arising from the electric fields in 
the molecule has been shown to be very small compared to the part 
having ``magnetic'' origin and can be neglected \cite{Bahloul98JPB}. 
Then, we split the {\bf C}-tensor arising from magnetic fields of 
the moving charges into two parts,  
\begin{equation}
{\bf C}={}^e{\bf C}+{}^n{\bf C},     
\label{split}
\end{equation}
produced by the molecular electrons and nuclei, 
respectively. The nuclear contribution, $^n$C, is obtained 
as a first order average in the vibrational ground state, and
depends only on the nuclear coordinates. Contrary, the electron 
part, $^e$C, is a second order series expansion which involves 
the full electronic spectrum. Fortunately, $xz$ and $zx$ components 
which are responsible for the mixing of states having $\Delta K = 1$
in C$_2$H$_4$ are vanishing by symmetry requirements \cite{Ilisca98PRA}.

Therefore, the {\bf C}-tensor, effective in ortho-para conversion, 
can be written (in Hz) for the $i$-th as \cite{Bahloul98JPB} 
\begin{eqnarray}
     {\text{C}}^{(i)} & = &{\sum_{k\neq i}} b_k\left[ 
          ({\bf r}_k\bullet {\bf R}_k){\bf 1} - 
           {\bf r}_k {\bf R}_k\right]\bullet {\bf B}; \nonumber \\
       b_k &=& 2 \mu_p q_k\big/ c \hbar R^3_k \ ,
\label{nc}
\end{eqnarray}
where {\bf R}$_k$ is the radial vector from the proton H$^{(i)}$
to the  charge $k$; {\bf r}$_k$ is the radial from the center of mass to 
the particle $k$; $q_k$ are the nuclei' charges; {\bf B} is the inverse 
matrix of inertia moment. {\bf B} is a diagonal matrix having the 
elements $B_{xx}= 58.6$~GHz, $B_{yy}=48.7$~GHz, and $B_{zz}=291.9$~GHz.
Index $k$ runs here over all nuclei in the molecule except the
proton $i$.

The spherical components of the spin-rotation
tensor of the rank $l$ ($l=1,2$) for the $i$-th proton calculated in 
the molecular frame, ${\cal C}^{(i)}_{l,q}$, can be determined using 
Eq.~(\ref{nc}). For the ethylene molecular structure from the 
Ref.~\cite{Volkenshtein72} and bare nuclei' charges these components are
\begin{eqnarray}
           &&  {\cal C}^{(1)}_{2,1}=3.8~{\text{kHz}};\ \   
             {\cal C}^{(1)}_{1,1}=-2.9~{\text{kHz}}; \nonumber \\
           &&   {\cal C}^{(3)}_{2,1}=-4.1~{\text{kHz}};\ \   
             {\cal C}^{(3)}_{1,1}=3.2~{\text{kHz}}.  
\label{c}
\end{eqnarray}
The difference between absolute values of ${\cal C}^{(1)}_{l,q}$ 
and ${\cal C}^{(3)}_{l,q}$
appears because of the shift of the molecular center of mass 
caused by the bigger mass of $^{13}$C in comparison with $^{12}$C. 
 
Similar to the previous section,
one can reduce the calculation of the matrix elements of the spin-rotation
coupling in asymmetric tops to the calculation of the symmetric-top
matrix elements. The latter can be found in 
\cite{Guskov95JETP,Ilisca98PRA,Guskov99JPB}. Thus one has the
the strength of mixing due to the spin-rotation coupling in
ethylene,
\begin{eqnarray}
  F_{SR}(a'|a)& = & 2(2J'+1)(2J+1)\sum_i\Bigg|
                   \sum_{K>0,q}A'_{K+q}A_K 
                    \Phi(i;J',K+q|J,K)  \nonumber \\
          &  & +\frac{1+(-1)^{J+p}}{\sqrt{2}}A'_1A_0
                \Phi(i;J',1|J,0)\Bigg|^2 .    
\label{fsr}        
\end{eqnarray} 
Here $q=\pm1$; $i=1,3$ denotes the hydrogen nuclei 1 and 3. Note,
that the protons 2 and 4 were also taken into account because they
produce mixing equal to that of protons 1 and 3. In (\ref{fsr}) 
the notation was used 
\begin{eqnarray}
    \Phi(i;J',K'|J,K) & = & \sum_l \sqrt{2l+1}\, {\cal C}^{(i)}_{l,q}
                  \left(\begin{array}{ccc}    
                      J' & l & J \\            
                     -K' & q & K               
                  \end{array}\right)  \times   \nonumber \\           
       &  &  \left[ y(J)(-1)^l \left\{\begin{array}{rcr}    
                                     J'& J & l \\            
                                     1 & 1 & J               
                               \end{array}\right\} + 
                  y(J') \left\{\begin{array}{rcr}    
                               J & J'& l \\            
                               1 & 1 & J'               
                               \end{array}\right\} \right],
\label{fjk}
\end{eqnarray} 
where \{:\ :\ :\} stands for the 6j-symbol; $y(J)=\sqrt{J(J+1)(2J+1)}$.   
The selection rules for the ortho-para mixing by spin-rotation 
perturbation in ethylene read
\begin{equation}
     \Delta p = 0;\ \ |\Delta J| \leq 1.
\label{selSR}
\end{equation}
And again, parity of ${\cal K}'$ and ${\cal K}$ is opposite.

\section{Conversion rates and discussion}

The decoherence rate, $\Gamma$, is another important parameter 
for the isomer conversion by quantum relaxation (see Eq.~(\ref{gamma})). 
This parameter is rather
difficult to calculate or to measure. Due to its physical 
meaning the value of $\Gamma$ is close to the rotational 
state population decay, which determines the line broadening
in microwave rotational spectra. Unfortunately, the ethylene does not have
pure rotational spectra because it has no permanent electric dipole moment.
Or, to be precise, very small dipole moment in the case of  
$^{13}$CCH$_{4}$. 

In the present calculations we take the estimation of $\Gamma$
from the line broadening in infrared spectra, for which the
information is available. Recently the line broadening was
accurately measured for the C$_2$H$_4$ molecules embedded in nitrogen
\cite{Blanquet00JMS}. One should not expect big difference
for the line broadening in the isotope species C$_2$H$_4$
and $^{13}$CCH$_{4}$. From the data \cite{Blanquet00JMS} one 
can estimate the decoherence rate as 
\begin{equation}
     \Gamma/P=2\cdot10^7~{\text{s}}^{-1}{\text{/Torr}}.
\label{decoh}
\end{equation}
This rate is 10 times smaller 
than the $\Gamma$ determined for the isomer conversion in CH$_3$F
\cite{Chap99ARPC}. The decrease of $\Gamma$ is because ethylene is not
a polar molecule and thus has its rotational relaxation slower. 
Probably, the estimation (\ref{decoh}) is too low 
for the case of pure ethylene gas because ethylene has polarizability 
larger than nitrogen. For the buffer gases having large polarizability, 
like Kr, or SF$_6$, nearly three times bigger line broadening 
was found \cite{Nagels96PRA53}. Nevertheless, we will 
use the estimation (\ref{decoh}) which can be considered as a low limit
for $\Gamma$.

\begin{figure}[htb]
\centerline{\psfig
{figure=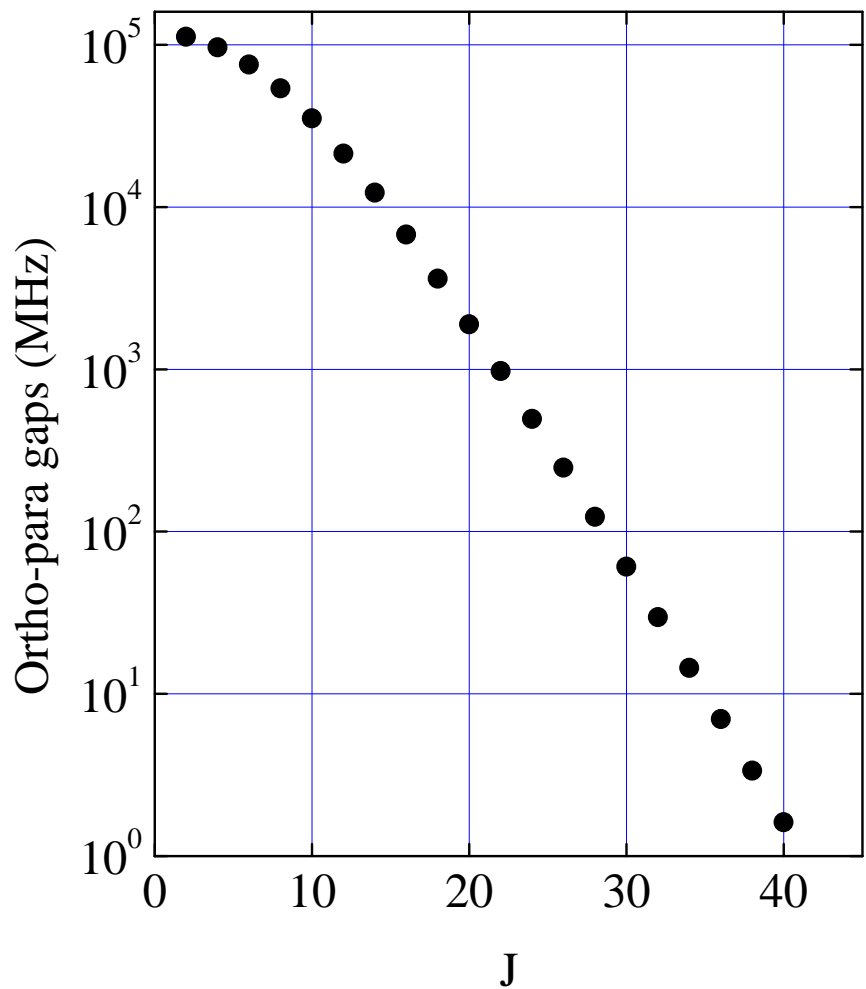,height=20cm}}
\vspace{0.5cm}
\caption{\sl Collapse of the ortho and para states in ethylene.
The figure shows the energy gaps between the ortho states 
($0,J,{\cal K}=0$) and the para states ($0,J,{\cal K}=1$).}
\label{fig4}
\end{figure}

\begin{figure}[htb]
\centerline{\psfig
{figure=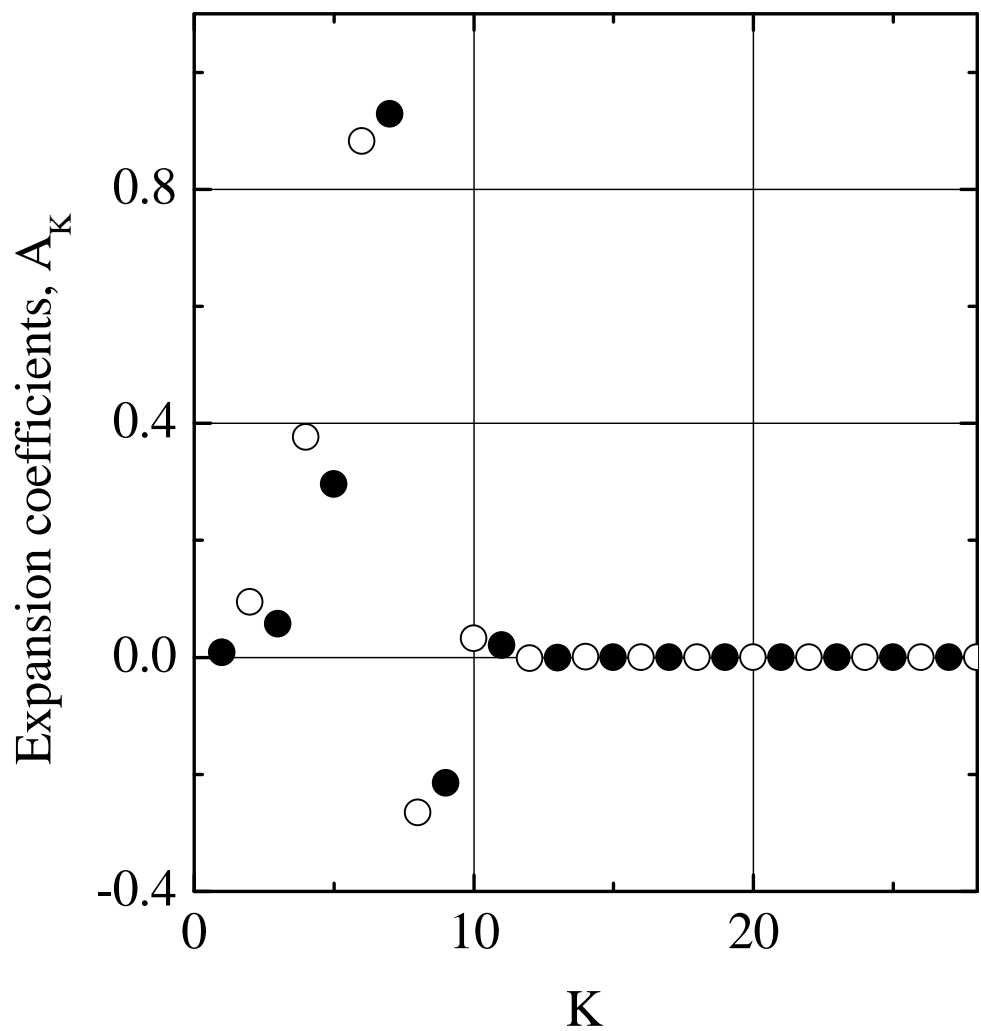,height=22cm}}
\vspace{0.5cm}
\caption{\sl The expansion coefficients, $A_K$, for the states 
most important for the spin conversion in $^{13}$CCH$_{4}$. 
(o)--ortho state ($1,J=28,{\cal K}=6$); 
($\bullet$)--para state ($1,J=27,{\cal K}=7$).}
\label{fig5}
\end{figure}

Now we are ready to calculate the conversion rate in ethylene. 
The final expression for the conversion rate, $\gamma$, is 
again the Eq.~(\ref{gamma}) having the strength of mixing,   
$F(a'|a)=F_{SS}(a'|a)+F_{SR}(a'|a)$. The results of the 
calculations are given in the Table~2. From these data we note 
that there is no appreciable contribution to the conversion from the 
collapsing ortho-para level pairs presented in Fig.~\ref{fig4}.
Another observation is that the contribution from the mixing of
states having $|\Delta{\cal K}|>1$ is also small. This is the
consequence of the fact that the ethylene molecule is rather 
close to a symmetric top. The main contributions to the conversion 
come from just two level pairs, both having
the same quantum numbers ($J'$,${\cal K}'$)--($J$,${\cal K}$), but 
different $p$. The most important ortho-para level pair
(1,27,7)--(1,28,6) has the frequency gap $\simeq$1.0~GHz. This level 
pair is indicated in Fig.~\ref{fig1} from which one can see that 
this pair of states is situated at rather high energies. 
Nearly 10\% of the conversion rate is due to the pair
(0,27,7)--(0,28,6). This level pair has the same strength of 
mixing as the first pair but three times bigger level splitting, $\omega$. 
The properties of the states ($1,J=28,{\cal K}=6$) and 
($1,J=27,{\cal K}=7$) are illustrated in Fig.~\ref{fig5} which 
presents the expansion coefficients, $A_K$, for these states. 
One can note that the biggest coefficient in both expansions
has the value of $K$ equal to the value of ${\cal K}$ chosen
for the designation of the eigenstate in our classification scheme. 

The total conversion rate in $^{13}$CCH$_{4}$ which is the sum over 
all ortho-para level pairs having $J'$ and $J$ up to 40, was found equal 
\begin{equation}
     \gamma/P = 2.7\cdot 10^{-4}~{\text{s}}^{-1}/{\text{Torr}}.
\label{gtheor}
\end{equation}
This rate is close to the experimental value 
$(5.2\pm0.8)\cdot 10^{-4}$~s$^{-1}$/Torr \cite{Chap00CPL}.  

The calculations of the conversion rates were repeated using less 
accurate molecular parameters from \cite{Vleeschouwer82JMS}. We 
obtained essentially the same results. The conversion is again 
determined by the same two level pairs and the value of the total 
conversion rate is $3.5\cdot 10^{-4}$~s$^{-1}$/Torr, which is close 
to the more precise value (\ref{gtheor}).

Our model allows to determine the pressure dependence of the 
conversion rate in $^{13}$CCH$_{4}$ at the conditions of the
experiment \cite{Chap00CPL}. By comparing the level splitting
of the most important ortho-para level pair (1,27,7)--(1,28,6), 
which is $\sim$1.0~GHz, with the decoherence
rate (\ref{decoh}) one can conclude that the case corresponds
to the limit $\Gamma\ll\omega$. From Eq.~(\ref{gamma}) one can see 
that in this limit linear grows of the conversion rate, $\gamma$,
versus gas pressure should take place. (Note that $\Gamma$ is 
proportional to the gas pressure.) This dependence was indeed 
observed in the experiment \cite{Chap00CPL}.

Magnitude of the calculated conversion rate relies 
heavily on the determination of the ortho-para
level splitting. The gap of the most important level pair appeared to be
rather large, $\simeq$1.0~GHz, and is unlikely that it has been determined 
with significant error. This can be proven by comparing this splitting
with the one calculated using less accurate molecular parameters
\cite{Vleeschouwer82JMS}. These parameters give the gap
0.914~GHz, which is close to the value given in the Table~2.

The two sources of the ortho-para mixing have been analysed. 
The spin-spin interaction between the molecular nuclei was 
possible to calculate 
rather accurately. The uncertainty of this perturbation is mainly
due to small errors in the knowledge of the molecular structure.
On the other hand, our analysis has shown that the 
spin-spin coupling contributes less than 1\% to the ethylene conversion 
and thus negligible. There are a few reasons for this reduction.
First, the spin-spin tensor has relatively small value due to the big
distances between interacting nuclei and small magnitude of the
magnetic moment of the carbon nucleus. Second, due to high symmetry 
of the molecular structure interactions between many pairs of protons
do not contribute to the ortho-para mixing. Moreover, there are no 
close ortho-para level pairs at small $J$, where spin-spin coupling
can compete with the spin-rotation coupling.

We have taken into account also the spin-rotation perturbation.  
The magnitude of the spin-rotation tensor was estimated and it was
demonstrated that the main contribution to the ethylene conversion 
comes from the spin-rotation mixing.  
Strong contribution from the spin-rotation coupling is due to the fact that 
the strength of the mixing $F_{SR}$ grows as $J^3$ in comparison with
the slower ($J^2$) grows of the spin-spin strength of mixing $F_{SS}$.
This makes the ``accidental'' resonance at big $J$ (1,27,7)--(1,28,6)
being so important. An extra cause of the efficiency of the spin-rotation
coupling is that the spin-rotation tensors of first and second 
rank contribute both to the conversion. On the other hand, not all close
ortho-para level pairs contribute to the conversion. In the particular case 
of the collapsing ortho-para levels (Fig.~\ref{fig4}) there is no strong
mixing mainly because of interferences between many symmetric-top 
components with different quantum numbers $K$, and also because for these
$J'=J$ pairs the first rank spin-rotation tensor does not contribute. 

Error in the estimation of $\Gamma$ introduces systematic uncertainty 
into the final result. But it does not undermine the model as a whole. 
First, one can turn around the approach
and determine which decoherence rate $\Gamma$ should be chosen
in order to fit the experimental data. In order to reproduce the 
experimental rate of conversion one should take the
decoherence rate, $\Gamma/P = 3.9\cdot 10^{7}$~s$^{-1}$/Torr, which is
less than two times different from our rough estimation (\ref{decoh}). 
Of course, such approach will be justified only when the spin-rotation 
perturbation in ethylene will be known from an independent source. 
There is another point which is worth to mention. Recently it was 
proposed a new approach to the isomer conversion in which the 
knowledge of $\Gamma$ is not necessary at all \cite{Chap00EPJD}. 
It is based on a fast linear sweep of the molecular levels through 
the ortho-para resonance. Nuclear spin conversion in such experimental
arrangement does not depend on $\Gamma$ and is dependent on the
strength of ortho-para mixing and the population of the mixed states.

Finally, it seems interesting to emphasize the difference in the 
conversion rates produced by the two pairs of identical rotational
quantum numbers (27,7)--(28,6) but different in parity. This peculiarity
arises from the asymmetric top properties, where energy of states
depend on parity. This in turn has dynamical consequence that the
spin isomers equilibrate first in one parity manifold (here the odd
one). 

\section{Conclusions}

The theoretical model for the nuclear spin conversion in ethylene
($^{13}$CCH$_{4}$) has been developed. For the first time a theory of
the spin isomer conversion in asymmetric tops was possible to
compare with experimental data on the isomer conversion.
We have found that the two experimental results
\cite{Chap00CPL}: the magnitude of the ethylene ($^{13}$CCH$_{4}$) 
isomer conversion 
rate and its pressure dependence, are consistent with the spin conversion
governed by quantum relaxation. The ortho-para state mixing is performed
in this molecule mainly by the coupling between the protons' spins
and the molecular rotation. We have identified also the two 
pairs of ortho-para states which are almost completely determine 
the spin conversion in $^{13}$CCH$_{4}$

\section*{Acknowledgments}

The authors are indebted to A.~Fayt for the possibility to use
the latest set of the ethylene molecular parameters prior to publication
and to M.~Irac-Astaud for stimulating discussions.

\newpage
Table~1. The character table for the symmetry groups C$_{2v}$, 
C$_{2v}$(M), D$_2$  and classification of the basis states (\ref{basis}).

\vspace{1cm}
\begin{tabular}{|c|cccc||ccc|}
\hline
C$_{2v}(M)$& E & (12)(34)  & E$^*$     & (12)(34)$^*$ & ortho & ortho & para  \\
\hline
C$_{2v}$ &E&C$_{2}$&$\sigma_v$&$\sigma_v'$ &&&  \\
\hline
D$_2$    & E & R$^\pi_z$ & R$^\pi_y$ & R$^\pi_x$ & $K$-even &$K$=0 & $K$-odd  \\
\hline                                                                                      
A$_{1}$  &1& 1  &     1    &   1       &$p$=0&$J,p$-even,&--        \\
B$_{2}$  &1&-1  &    -1    &   1       &--        &--                  &$p$=1\\
A$_{2}$  &1& 1  &    -1    &  -1       &$p$=1&$J,p$-odd, &--        \\
B$_{1}$  &1&-1  &     1    &  -1       &--        &--              &$p$=0\\
\hline
\end{tabular}

\newpage
Table~2. The most important ortho-para levels and their contributions
to the spin conversion in ethylene. In calculations the molecular parameters
from \cite{Fayt99} were used. The total rate combines the contribution
from all ortho-para level pairs having $J\leq40$.

\vspace{1cm}
\begin{tabular}{|cccccc|}
\hline
 Level pair & Energy & $\omega/2\pi$ & $F_{SS}$ & $F_{SR}$ & $\gamma/P$ \\
 $p',J',{\cal K}'$-$p,J,{\cal K}$ & (cm$^{-1}$) &(MHz)        
 & (MHz$^2$) & (MHz$^2$) & (10$^{-4}$~s$^{-1}$/Torr) \\
\hline
1,27,7--1,28,6 & 871.53 & 1006.7 & 4.2$\cdot10^{-2}$ & 4.4 & 2.49  \\
0,27,7--0,28,6 & 871.69 &-3435.5 & 4.2$\cdot10^{-2}$ & 4.4 & 0.21  \\
0,4,3--0,6,2   &  53.72 &-3180.6 & 1.1$\cdot10^{-3}$ & 0   & 3.4$\cdot10^{-3}$  \\
1,26,3--1,24,6 & 680.68 &  944.3 & 3.4$\cdot10^{-4}$ & 0   & 5.6$\cdot10^{-4}$  \\
0,20,1--0,20,0 & 356.02 & 1894.4 & 7.3$\cdot10^{-8}$ & 2.4$\cdot10^{-6}$ 
                                                       & 4.8$\cdot10^{-6}$ \\
\hline
Total rate &&&&& 2.70 \\
\hline
\end{tabular}

\newpage

\newpage

\newpage

\newpage

\newpage

\end{document}